\documentclass[conference]{IEEEtran}
\IEEEoverridecommandlockouts

\usepackage{cite}
\usepackage{amsmath,amssymb,amsfonts}
\usepackage{algorithmic}
\usepackage{graphicx}
\usepackage{textcomp}
\usepackage{xcolor}
\usepackage[hyphens]{url}
\usepackage{braket}
\usepackage{subcaption}
\usepackage{array}
\usepackage{comment}
\usepackage{tabularx}

\def\BibTeX{{\rm B\kern-.05em{\sc i\kern-.025em b}\kern-.08em
    T\kern-.1667em\lower.7ex\hbox{E}\kern-.125emX}}
\begin{document}

\pdfpagewidth=8.5in
\pdfpageheight=11in


\pagenumbering{arabic}

\title{Heterogeneously error-corrected QRAMs}
\author{\IEEEauthorblockN{Ansh Singal}
        \IEEEauthorblockA{\textit{Electrical Engineering} \\
        \textit{Northwestern University}\\ Evanston, USA \\ asingal@u.northwestern.edu}
        \and
        \IEEEauthorblockN{Kaitlin N. Smith} \IEEEauthorblockA{\textit{Computer Science} \\ \textit{Northwestern University}\\ Evanston, USA \\ kns@northwestern.edu}
    }

\maketitle
\thispagestyle{plain}
\pagestyle{plain}


\begin{abstract}
        Quantum Random Access Memory (QRAM) holds the promise of enabling several large scale applications of quantum computers. However, designing fault tolerant QRAMs for large scale applications is still an open problem due to the poor error and resource scaling of current architectures. Existing protocols often overlook the need for error correcting QRAMs, which will be required for data-intensive, fault-tolerant applications. However, naively error correcting all qubits used to implement the QRAM is prohibitively resource intensive, quickly becoming infeasible for large applications. To fill this gap, we propose a novel QRAM architecture that leverages variable strength error correction. We strongly error-correct qubits that heavily influence query fidelity, and lightly correct less critical regions of the QRAM. This scheme produces queries with fidelity bounded by a constant for arbitrarily sized QRAMs without requiring improvements in physical hardware. Furthermore, the heterogeneous scheme requires 5x fewer resources (for depth 30 QRAM) and quadratically slower error scaling as compared to a uniformly error corrected Bucket Brigade QRAM. In this work, we present a rigorous analysis of the query fidelity scaling and perform resource analyses of two variations of the heterogeneous architecture using the surface code. We verify our results using numerical simulations and compare our results against several other existing QRAM techniques. Through our results, we quantitatively prove the optimal scaling of the heterogeneous architecture, paving a way for data-intensive and fault tolerant quantum applications. 

 
\end{abstract}
    
    \section{Introduction}
        Quantum computing is at a pivotal moment where error rates are scaling below the thresholds required for demonstrations of logical qubits~\cite{google2025quantum}. While quantum processors continue to improve in capacity and performance, unanswered questions remain about the architectures needed to support the powerful quantum algorithms capable of addressing problems that are currently intractable. For example, QRAMs are required for efficient and coherent access to databases, either quantum or classical, in many quantum algorithms. This algorithmic building block is essential for solving problems that compute using external data, and example applications that leverage QRAM include unstructured database search \cite{grover1996fast}, Hamiltonian simulation \cite{childs2018toward} and Quantum Machine Learning \cite{wang2024comprehensive}. Unlike a classical RAM, where the memory returns one unit of data per address query ($\ket{i}\xrightarrow{\text{RAM}}\ket{\psi_i}$), a QRAM can access data in arbitrary superpositions, thereby performing the following unitary operation: 
    
        \begin{equation}
            \sum_{i=0}^{N-1}\alpha_i\ket{i}_A\ket{0}_B\xrightarrow{\text{QRAM}} \sum_{i=0}^{N-1}\alpha_i\ket{i}_A\ket{\psi_i}_B.
        \end{equation}
    
        \noindent 
        In the rest of the paper, we represent the QRAM capacity, or the number of entries it stores, as $N$ and $n =\log_2(N)$.
        
        The flexibility and power of the QRAM come at two costs ubiquitous in quantum computing: large resource overhead of implementing QRAMs and decoherence during long algorithm executions \cite{arunachalam2015robustness}. Although modern QRAM architectures are designed to be inherently error resilient (e.g. \cite{giovannetti2008architectures}), algorithms which require large QRAMs, and several data accesses with those QRAMS, require high fidelity queries, which motivates the use of quantum error correction (QEC). However, actively error-corrected QRAMs require significant physical qubit overheads~\cite{di2020fault}, underscoring the need for new approaches to make QRAM practical.
        
        Most proposals for QRAM devices require physical component fidelities to improve as the size of the QRAM increases \cite{dalzell2025distillation}. This bottlenecks how far QRAM devices can be scaled given error-limited physical components. Approaches that distill a QRAM resource state to apply the QRAM operation fault-tolerantly suffer from such scaling constraints. Hence, literature lacks a clear path to efficiently scalable fault-tolerant QRAMs, and this could deter the use of quantum computers for data intensive applications.

        In this work, we propose, develop, and analyze a novel architecture for implementing scalable QRAM devices which do not suffer from the drawbacks of contemporary QRAM designs. Our Heterogeneous QRAM architecture utilizes the internal structure of the Bucket Brigade QRAM to efficiently error-correct queries, thereby paving a way to scalable quantum quantum computing using quality constrained hardware. The main contributions of this paper are as follows:
        
        \begin{enumerate}
            \item We propose a QRAM architecture which produces queries independent of the size of the QRAM without requiring strict hardware requirements that other approaches demand for larger QRAMs.  
            \item We propose two implementations for the heterogeneous scheme and perform error and resource scaling studies for each implementation. 
            \item We verify the error scalings for the implementations using numerical simulation and demonstrate over $5\times$ reduction in resources required to implement the QRAM (as compared to uniformly error corrected QRAMs). 
        \end{enumerate}
        
        The rest of the paper is organized as follows: after a survey of previously proposed QRAM architectures (Sec. \ref{sec:QRAM_architectures}) and an overview of our chosen QEC scheme, the surface code (Sec.~\ref{sec:surface_codes}), we present the heterogeneous architecture (Sec. \ref{sec:architecture}) and an error analysis (Sec. \ref{sec:error_analysis}). We present two implementations of the QRAM: Fat Tree inspired (FT-Hetero, Sec.\ref{sec:implementation}) and Bucket Brigade inspired (BB-Hetero, \ref{sec:alternative_implementation}). We verify our analytical results numerically in Sec. \ref{sec:numerical_sim}. In Sec. \ref{sec:comparison}, we compare the the heterogeneous architecture against other QRAM architectures. Finally, we conclude the paper in Sec. \ref{sec:conclusion}. 
        
    \section{QRAM Architectures} \label{sec:QRAM_architectures}
        Several QRAM architectures have been proposed in literature that explore the tradeoffs between query latency, qubit overhead, and resilience to QRAM decoherence. Sequential query circuit (SQC) based architectures like the one proposed in \cite{babbush2018encoding} focus on qubit efficiency of the QRAM and require $O(n)$ qubits for its implementation. This qubit efficiency, however, comes at the cost of query depth, requiring $O(N)$ operation depth per query. As a note, the architecture proposed in \cite{babbush2018encoding} is a QROM architecture, which can read a classical database but cannot wite back to the database. SQC architectures, however, can be enabled to both read and write to the database. 
        
        Router-based architectures such as the Fanout Memory \cite{nielsen2010quantum} and the Bucket Brigade (BB) QRAM \cite{giovannetti2008quantum} aim to reduce query time and require $O(n)$ operation depth per query. However, these architectures require $O(N)$ qubits to implement the QRAMs. Other proposals, like the SELECT-SWAP QRAM \cite{low2024trading}, aim to optimize $T$-gate counts of QRAM query circuits by utilizing a larger number of qubits while utilizing unused ``dirty" qubits to supplement the larger qubit overhead. This is possible because the QRAM is indifferent to the initial state of a subset of its qubits.  
        
        Several recent works have focused on hybrid architectures, such as \cite{xu2023systems} which combines SQC and BB architectures, along with other optimizations, to reduce the $T$-depth of the BB QRAM to orders similar to the SELECT-SWAP QRAM. Furthermore, \cite{zhu2024unified} combines the SELECT-SWAP and BB QRAMs to reduce query infidelity of the raw SELECT-SWAP QRAM.  
        
        All the QRAMs discussed so far use the same circuit for querying any classical database. The physical qubits corresponding to the classical data can be modified to query a different database. However, databases with skewed data (i.e. mostly zeros or ones) can be queried efficiently by designing special query quantum circuits based on the data in the database \cite{di2020fault}. However, such QRAMs that leverage knowledge of the database prove inefficient in terms of compilation overheads in the case that the database values are frequently updated. 
        
        \subsection{Bucket Brigade QRAM} \label{sec:BB_QRAM}
            The architecture proposed in this work is based on the Bucket Brigade (BB) QRAM \cite{giovannetti2008quantum}, and hence, we take a deeper look at this architecture. The BB QRAM is a router-based QRAM architecture where routers are entangled with address bits in a $W$-like state\cite{hann2021resilience} to query a database in a coherent superposition. The original BB QRAM proposal used qutrits for holding router data. The three states were $\ket{0}, \ket{1}$ and $\ket{W}$. All the routers start off in the stable $\ket{W}$ (wait) state. When the router is in the wait state, it does not perform either of the routing operations. The other two states, $\ket{0}$ and $\ket{1}$, are responsible for performing the routing operations to the left and right sub-trees respectively.  
            
            The resilience of the BB QRAM was studied in \cite{hann2021resilience} which showed that the BB QRAM has favorable query infidelity scaling due to the low entanglement between the routers. 
            Furthermore, the tree-like structure of the BB QRAM ensures that the routers at level $l$ (where the root is at $l=0$) are included only in $N2^{-l}$ branches. Hence, the infidelity only improves with active error correction, where the logical infidelity scales as:
            \begin{equation}
                1-F_L \leq 4\epsilon_L T_L\log N
                \label{eqn:original_infidelity}
            \end{equation}
            where $F_L$ is the query fidelity of the QRAM, $\epsilon_L$ is the logical error rate of the qubits in the QRAM and $T_L$ is the time required to perform the query. 
    
    \section{Surface Code Error Correction} \label{sec:surface_codes}
        As we will see in Sec. \ref{sec:architecture}, our architecture leans favorably towards any topological error correction code with exponential error suppression, but for simplicity of analysis, we consider a surface code implementation in this work. However, we note that the heterogeneous architecture is not limited to just surface code QEC. In this section, we provide an overview of the properties of the surface code that enable the performance advantages of the heterogeneous architecture.
        
        Surface codes are a promising QEC code for scalable fault tolerant quantum computation on a 2D lattice. Surface codes owe their recent rise in popularity to low connectivity constraints (i.e. nearest neighbor connectivity) and high physical error thresholds~\cite{fowler2012surface}. Additionally, the surface code patches can be deformed, making it adaptable to specific types of problems \cite{resch2022variable} or device configurations~\cite{lin2024codesign}. This makes the surface code favorable for encoding logical qubits on several physical qubit platforms like superconducting circuits \cite{google2025quantum} and neutral atoms \cite{bluvstein2024logical}. Recent breakthrough demonstrations such as \cite{google2025quantum} have shown that surface codes can be realized to enable scalable, fault-tolerant quantum computing. Lattice surgery \cite{horsman2012surface} is by far the most promising method for quantum computation based on the surface code, owing to its higher qubit and non-Clifford gate efficiency \cite{fowler2018low} as compared to other techniques such as defect-based computation. 
        
        The logical error rate for the surface code is defined as \cite{fowler2012surface}:
        \begin{equation}
            \epsilon_L = \epsilon'(p')^{d_e}
            \label{eqn:error_corrected}
        \end{equation}
        where $p' = p/p_\text{th}$, where $p$ is the physical error rate, $p_\text{th}$ is the surface code threshold error rate, $d_e=(d+1)/2$ if $d$ is odd and $d_e=d/2$ if $d$ is even, where $d$ is the code distance and $\epsilon'$ is an empirically derived constant ($\epsilon'=0.03$ in \cite{fowler2012surface}). Recent advancements in surface code quantum computing like magic state cultivation \cite{gidney2024magic} have made universal fault tolerant quantum computing using the surface code more efficient and brings large scale quantum systems closer to reality. 
        
        Variable strength quantum error correction \cite{resch2022variable} is a method of using QEC codes with variable code distances to decrease the resource overheads of error correction. Conventionally, it's used to correct portions of computers with high physical error rate more robustly as compared to less error-prone portions \cite{yin2024surf}. Lattice surgery enables efficient code deformation and variable code distance computing. 
        Factors like easily variable and manipulatable code distance, high error threshold and efficient magic state generation make the surface code a logical choice for implementing the heterogeneous architecture.

    \section{Heterogeneous QRAM architecture} 
        \subsection{Architecture}\label{sec:architecture}
            As seen in Sec. \ref{sec:BB_QRAM}, the number of branches in which a router at level $l$ is present scales as $N2^{-l}$. This provides the insight that uniformly error correcting all router qubits is wasteful and unnecessary. This is because errors near the root of the tree are exponentially more detrimental to query fidelity as compared to errors near the leaves of the tree. Hence, the routers near the root of the tree must have exponentially lower error rates as compared to error rates near the leaves of the tree. As noted in Sec. \ref{sec:surface_codes}, surface codes give us considerable flexibility in varying the logical error rates by varying the code distance of the code. Apart from improving the error scaling, the varying code distance restricts the qubit overhead in the heterogeneous scheme since as the number of routers increases near the leaves of the tree, the code distance decreases.    
            
\begin{figure}
                \centering
                \includegraphics[width=\linewidth]{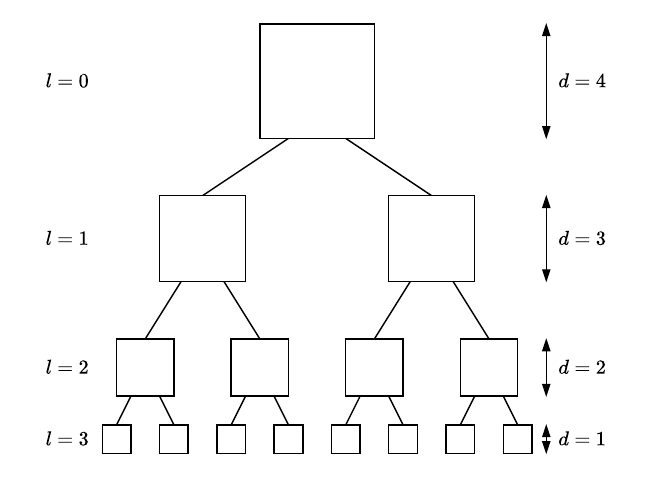}
                \caption{Heterogeneous QRAM architecture.}
                \label{fig:qram_arch}
            \end{figure}

            The routers near the root need to be more error resilient as compared to the routers near the leaves of the tree, so we progressively increase the code distance for encoding logical router qubits as we traverse up the tree from the leaves. Fig. \ref{fig:qram_arch} shows the architecture of such a QRAM with $N=8$. The code distance of each level is distributed by a function $f(l)$ where $l$ is the level of the tree (the root node is at $l=0$). In the example shown in Fig. \ref{fig:qram_arch}, the code distances are increased linearly, such that the code distance at level $l$ is $f(l) = n-l+1$. While other distributions which scale either faster (e.g. $f(l) = (n-l+1)^k$ where $k>1$) or slower (e.g. $f(l) = (n-l+1)^k$ where $k<1$) alter the error or resource properties of the QRAM, we focus on the case of linear distribution of code due to its simplicity and efficiency. 

        \subsection{Error Analysis} \label{sec:error_analysis}
            We perform an error analysis of the heterogeneous QRAM architecture. Our error analysis is based on the analysis performed in \cite{hann2021resilience} which we briefly revisit and extend it to the heterogeneous architecture. The fidelity of the QRAM in \cite{hann2021resilience} is 
            
            \begin{equation}
                \begin{split}
                    \mathbb{E}[F(c)]& \geq [2\mathbb{E}_n[\Lambda(c)]-1]^2,~~~~ \Lambda(c)\geq 1/2
                \end{split}
                \label{eqn:fidelity_eqn}
            \end{equation}
            where $c$ is an error configuration that specifies which Kraus operator is applied to each router at each time step. $\Lambda(c)$ is the weighted fraction of branches in configuration $c$ that are error free (i.e. branches that lie in the set of good branches, $g(c)$). To calculate the expectation of query fidelity $\mathbb{E}[F]$, we need the expectation value of $\Lambda(c)$ ($=\mathbb{E}_n[\Lambda(c)]$), which we can plug in Eqn. \ref{eqn:fidelity_eqn} to calculate $\mathbb{E}[F]$. We can calculate $\mathbb{E}_n[\Lambda(c)]$ recursively as \cite{hann2021resilience}:
            \begin{equation}
                \mathbb{E}_{i+1}[\Lambda] = (1-\epsilon_{i+1})^{T_{i+1}}\mathbb{E}_{i}[\Lambda]
                \label{eqn:recurrecnce}
            \end{equation}
            where $\epsilon_i$ is the per step error rate of the routers at the $i^{th}$ level, and $T_i$ is the query execution time at level $i$, in terms of number of steps. This means that routers at level $i$ need to be coherent for $T_i$ time for performing the query. If we keep $\epsilon_i = \epsilon$ and $T_i = T$ where all the qubits have the same error rate $\epsilon$ and all the qubits need to remain coherent for the entire duration of the query $T$, we get
            \begin{equation}
                \mathbb{E}_n[\Lambda] = (1-\epsilon)^{nT}
                \label{eqn:expected_good_c}
            \end{equation}
            which reproduces the result from \cite{hann2021resilience}, restated in Eqn. \ref{eqn:original_infidelity}.  
            
            Our approach differs from the above approach in that the qubits composing the QRAM are now error corrected. Furthermore, logical error rates of router qubits are not the same and are dependent on where the router is placed in the tree. Hence, for the heterogeneous architecture, qubit error rates are determined by $\epsilon_{i} = \epsilon'(p')^{d_e}$ (where $d_e = \lceil \frac{n-i+1}{2}\rceil$), based on the surface code error rate \cite{fowler2012surface}. We can then write $\mathbb{E}_n[\Lambda]$ for the heterogeneous architecture as
            \begin{equation}
                \begin{split}
                    \mathbb{E}_{n}[\Lambda] &= \prod^{n}_{i=0}(1-\epsilon_i)^{T_i}\\
                    &= \prod^{n+1}_{d=1}(1-\epsilon'(p')^{d_e})^{T_d}\\
                \end{split} 
                \label{eqn:expected_good_branches}
            \end{equation}
            where, for the second equality, we are iterating not by the level index ($i$), but the code distance ($d = n-i+1$), which goes from $d=1$ for the leaves to $d=n+1$ for the root node. 
            
            To evaluate this expression, we look at every term in the product as a first order approximation of the binomial term $(1-\epsilon')^{(p')^{d_e}}$. This is a fair approximation if $1\ll\epsilon'^2(p')^{d_e}$, which is true for empirical values of $\epsilon'$ ($\epsilon' = 0.03$ in \cite{fowler2012surface}) and realistic $p'$. Under this assumption, we write $\mathbb{E}_{n}[\Lambda(c)]$ as
            \begin{equation}
                \begin{split}
                    \mathbb{E}_{n}[\Lambda] &\approx \prod^{n+1}_{d=1}(1-\epsilon')^{(p')^{d_e}T_d}\\
                    &= (1-\epsilon')^{K(n)}\\
                \end{split} 
                \label{eqn:approximated_expected_branches}
            \end{equation}
            where 
            \begin{equation}
                K(n) = \sum^{n+1}_{d=1}(p')^{d_e}T_d .
                \label{eqn:K}
            \end{equation}
            
            \noindent Similar to \cite{hann2021resilience}, we can now substitute Eqn. \ref{eqn:approximated_expected_branches} in Eqn. \ref{eqn:fidelity_eqn} and use the Bernoulli inequality and again, which restricts our analysis to the leading order of error $\epsilon_d$. With this restriction, we find the following upper bound on the query infidelity as
            \begin{equation}
                \begin{split}
                    1-F(c)\leq 4\epsilon'K(n).
                \end{split} 
                \label{eqn:hetero_fidelity}
            \end{equation}

            \noindent Hence, the infidelity scaling for the heterogeneous QRAM depends on the $K(n)$ factor, which in turn, depends on the distribution of the error rates across the layers of the QRAM, and the time that the qubits in those layers need to remain coherent for. 

            In the original BB QRAM, the addresses are set in the tree first, and the bus is routed through the tree next. This requires all the qubits to remain coherent for $O(n)$ time (i.e. asymptotically similar amount of time as the entire query). This would reduce the potential advantage observed from the heterogeneous architecture since the low code distance routers would also have to wait for large code distance operations to execute, incurring lots of errors. Hence, for optimal performance, we need qubits with code distance $d$ to remain coherent only for a period of $O(\text{Poly}(d))$. If this is satisfied, we get (to leading order)
            \begin{equation}
                K(n) = \sum_{d=1}^{n+1}(p')^{d_e}d^t, ~~~~t>0
                \label{eqn:theoretical_K}
            \end{equation}
            which, is a constant, $K$. Substituting this into Eqn. \ref{eqn:hetero_fidelity} the query infidelity would scale as
            \begin{equation}
                1-F \leq 4\epsilon'K,
                \label{eqn:theoretical_infidelity}
            \end{equation}
            
            \noindent implying the query infidelity is upper bounded by a constant. Hence, for large-sized QRAMs, the query infidelity is independent of the size of the QRAM. In the following sections, we discuss possible implementations of the heterogeneous scheme for achieving this error scaling. 
    
        \subsection{Heterogeneous Implementation 1} \label{sec:implementation}
            In the previous section, we saw that the heterogeneous QRAM can theoretically produce constant infidelity queries for larger sized QRAMs. We now discuss a potential implementation for this QRAM. As discussed in Sec. \ref{sec:error_analysis}, most BB QRAM implementations employ a three step query process: address setting, bus routing, and retrieval (or uncomputation) \cite{xu2023systems}. Since the bus-routing step itself is $O(n)$, performing the query in such discrete steps would nullify all advantage of the heterogeneous scheme. So, we include the bus routing step within the address setting, to avoid the $O(n)$ coherence time overhead on all the routers. Furthermore, an operation with code distance $d$ requires $d$ code cycles for successful error decoding. So, any operation performed at the root node would require $n$ code cycles, which requires $O(n)$ time. We want to perform all large code distance operations before performing lower code distance operations, and this can be accomplished by routing the entire address register through the QRAM simultaneously. As we will see, this is like the error-rate analogue of the Fat Tree architecture for classical RAMs and hence, we call this the \textbf{Fat Tree inspired} implementation of the heterogeneous QRAM (FT-Hetero). 

            \subsubsection{Quantum Circuit}
\begin{figure}
                \includegraphics[width=0.5\textwidth]{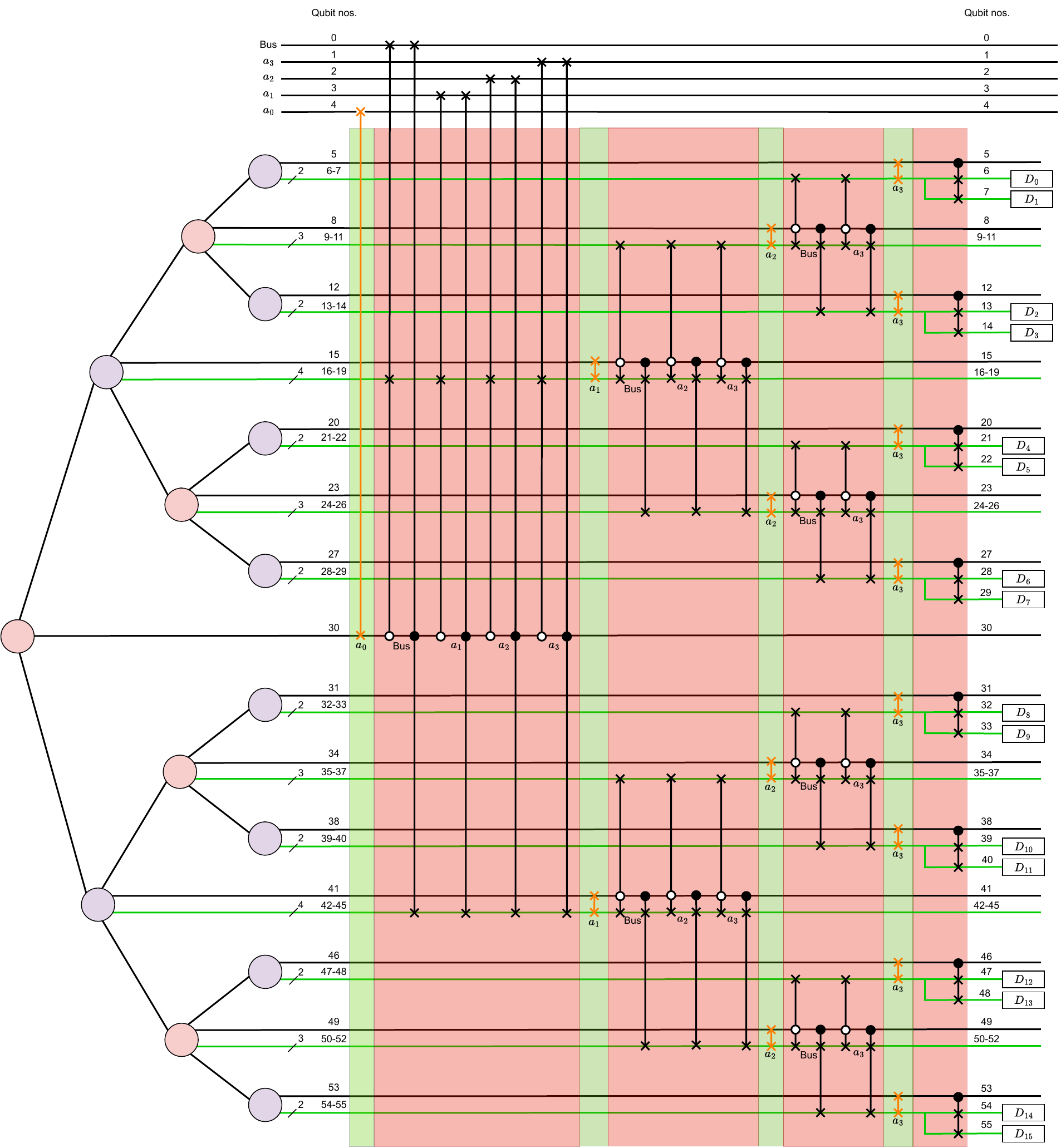}
                \caption{Query circuit for FT-Hetero implementation. Each router only needs $O(\text{Poly}(d))$ bus/address routing steps (red) and address setting steps (green), thereby producing optimal query fidelity scaling.}
                \label{fig:scheme2_circuit}
            \end{figure}

            Fig. \ref{fig:scheme2_circuit} shows the quantum circuit for the FT-Hetero architecture. The red sections correspond to the address + bus routing steps, and the green sections correspond to address setting steps. Each time an address bit is set, that bit is no longer router further in the tree. Hence, the number of qubits per node decrease as we go down the tree, as is visible in the register sizes marked in the figure. This is similar to the Fat Tree QRAM architecture where the bandwidth decreases as you traverse the tree. Since routers at level $i$ are involved in operations with code distance $d=n-i+1$ or lower, there are only $O(\text{Poly}(d))$ such operations. Hence, this architecture follows eqn. \ref{eqn:theoretical_infidelity}, thereby upper bounding the infidelity of our QRAM queries. 

            \subsubsection{Error Analysis}
            We can now determine Eqn. \ref{eqn:theoretical_K} for this approach. First, we calculate the coherence time required for routers in any level $j$
            \begin{equation}
                T_j = \sum_{i=j}^{n-1}[(n-i+2)\left[2(n-i+1)c\right]+(n-i+1)s],
                \label{eqn:scheme2_Tj}
            \end{equation}
            where $(n-i+1)s$ and $(n-i+1)c$ are the time for performing SWAP and CSWAP operations over $n-i+1$ code distance surface codes. We can now calculate Eqn. \ref{eqn:K} using Eqn. \ref{eqn:scheme2_Tj}. However, to calculate the expected fraction of branches that do not have an error, we cannot use Eqn. \ref{eqn:approximated_expected_branches} directly since each router does not have only one router qubit but the entire sets of address qubits. As a result, if any of the address bits is corrupted, at any level, its entire subtree will then be corrupted since the corrupted address qubit would propagate downwards, eventually settling at all of the routers corresponding to the corrupted address bit. The expected number of error-free branches depends not only on the set address qubit, but all the address qubits being error free. Using this, the fraction of qubits which are error free is:
            \begin{equation}
                \mathbb{E}_i[\Lambda] = (1-\epsilon')^{(n-i)T_{n-i+1}}\mathbb{E}_{i-1}[\Lambda],
            \end{equation}
            where $(n-i)$ is the number of address qubits at level $i$. Finding $\mathbb{E}_n[\Lambda]$ similarly to Eqn. \ref{eqn:expected_good_branches}, and making the approximation in Eqn. \ref{eqn:approximated_expected_branches}, we get an upper bound on infidelity as
            \begin{equation}
                    \begin{split}
                        1-F &\leq 4\epsilon'\sum_{d=1}^{n+1}p'^{d_e}(d-1)\sum_{d'=2}^{d}(d'+1)\left[2d'c+s\right]\\
                        &\leq 4\epsilon'K',
                    \end{split}
                \label{eqn:scheme2_error}
            \end{equation}
            \begin{equation}
                \begin{split}
                    &\text{where,}~K'= s\left[p'-\frac{p'^{3/2}}{(\sqrt{p}-1)^3}(p'+3p'^{1/2}+4)\right]+\\
                    &(s+2c)\left[2p'-\frac{p'^{3/2}}{(\sqrt{p}-1)^4}(2p'^{3/2}+8p'+13p'^{1/2}-10)\right]+\\
                    &2c\left[4p'-\frac{p'^{3/2}}{(\sqrt{p}-1)^5}(4p'^2-20p'^{3/2}+41p'+43p'^{1/2}+26)\right].\\
                \end{split}
            \end{equation}
            
            \noindent We observe that the infidelity is upper bounded by a constant, making the query infidelity independent of the size of the QRAM for larger QRAM sizes. Note that this query infidelity corresponds to the QRAM architecture with qutrit routers. This can be implemented using two error corrected qubits at each router rather than one qutrit. If we implement the QRAM with routers made of a single qubit instead, the query infidelity scales as $O(n)$. Appendix \ref{apdx:qubit_error_analysis} includes more analysis and proof of this error scaling.  

            \subsubsection{Qubit Overhead}
            We can find the approximate number of physical qubits required to implement FT-Hetero. We use the packing efficiency for the surface code using lattice surgery in \cite{fowler2018low}: $3d^2$ physical qubits per logical qubit with code distance $d$. So, the number of qubits vary by the layer as:
            \begin{itemize}
                \item Layer $0$: Logical qubits: 6, $d=5$ 
                \item Layer $1$: Logical qubits: 10, $d=4$ 
                \item Layer $2$: Logical qubits: 16, $d=3$ 
                \item Layer $i$: Logical qubits: $2^i(n-i+2)$, $d=n-i+1$ 
            \end{itemize}
            \begin{equation}
                \therefore M_{\text{total}} = 3\sum_{i=0}^{n}2^i(n-i+2)(n-i+1)^2.
            \end{equation}
            Setting $n\to\infty$, $M_{\text{total}} = 192N$. This shows that asymptotically, the number of physical qubits grows similar to the uncorrected QRAM circuit, albeit with a large constant factor. We improve the constant factor slightly to $153N$ in Appendix \ref{apdx:efficeint_resources}. This constant factor is attributed primarily to the poor resource scaling of the surface code. Since our architecture is amicable to any topological code which demonstrates exponential error suppression by code distance, more efficient error correction codes could be used to implement FT-Hetero more efficiently.       
            
        \subsection{Heterogeneous Implementation 2} \label{sec:alternative_implementation}
            Although the implementation in Sec. \ref{sec:implementation} produces queries with infidelity upper bounded by a constant, it also has a few drawbacks. First, and the most obvious, is the large qubit overhead. Although it has the same asymptotic scaling as the physical BB QRAM, it has a large constant factor. Also, since the implementation required routing the entire address register simultaneously, the operation depth per query scaled as $O(n^2)$. Although it did not affect the query fidelity, each query took longer to execute. In this section, we explore an alternative heterogeneous architecture that improves on these factors. 
    
            We can reduce the number of qubits and time per query using a query circuit similar to that in \cite{hann2021resilience} with a modification where the bus qubit is routed along with the address qubits for the same reason described in Sec. \ref{sec:implementation}, but the address qubits are routed sequentially and not simultaneously. Hence, the routers do not need to hold the entire address register, reducing the qubit overhead of the QRAM while also reducing the query time. Since this is similar to the BB circuit in \cite{hann2021resilience}, we call this the \textbf{BB-inspired} Heterogeneous implementation. 
    
            \subsubsection{Quantum Circuit}
\begin{figure*}
            \centering
            \begin{subfigure}{0.6\linewidth}
                \includegraphics[width=\textwidth]{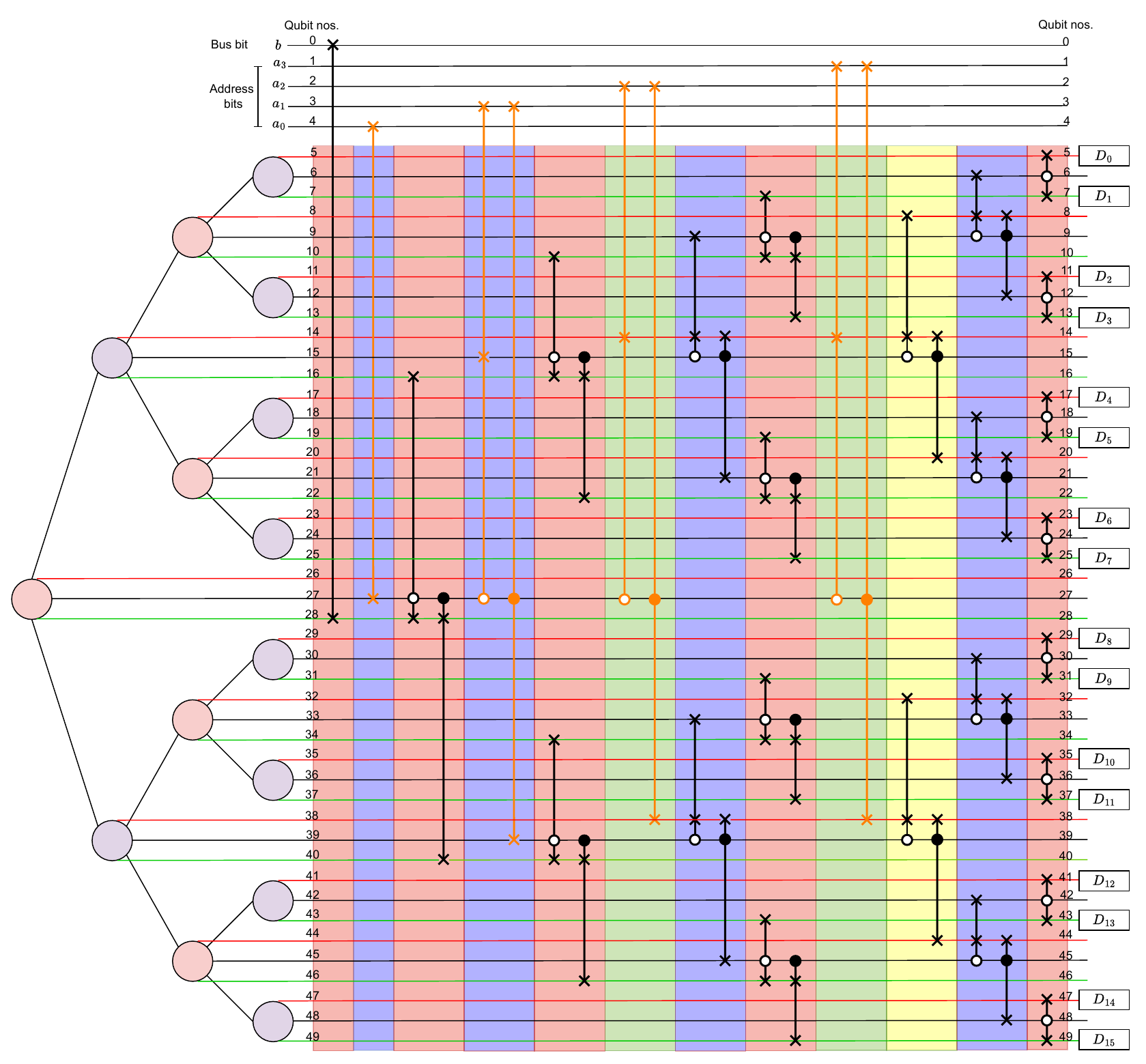}
                \label{fig:query_final_circuit_N_8}
                \caption{Query circuit ($N=8$)}
            \end{subfigure}~~~~~~~~
            \centering
            \begin{subfigure}{0.35\linewidth}
                \includegraphics[width=\textwidth]{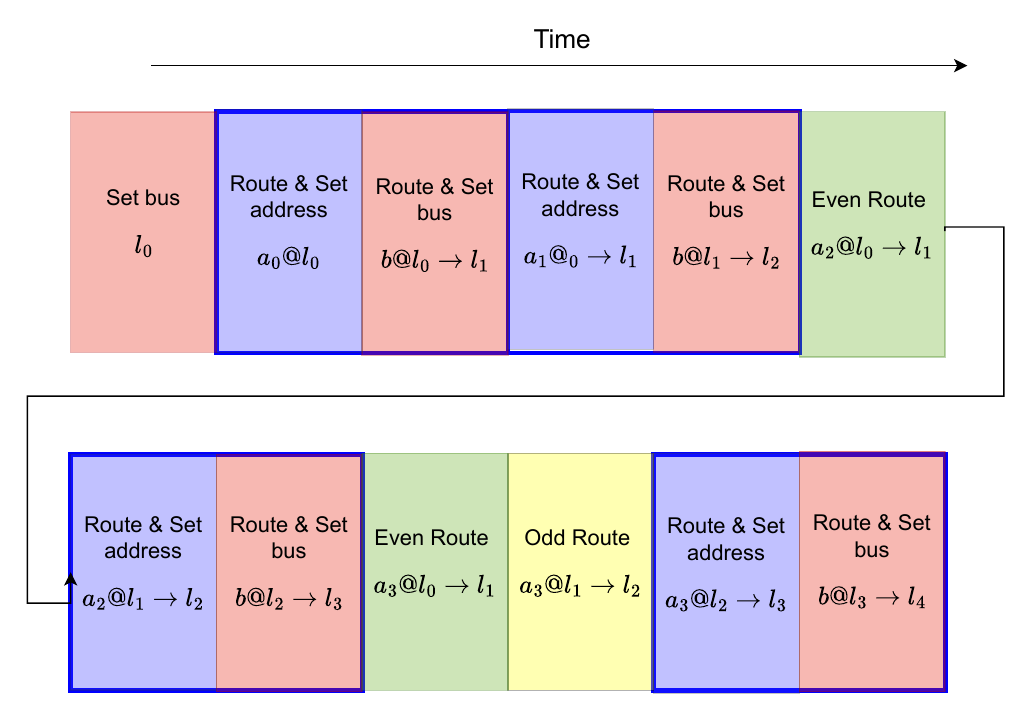}
                \label{fig:query_steps_N_8}
            \\\\
                \includegraphics[width=\textwidth]{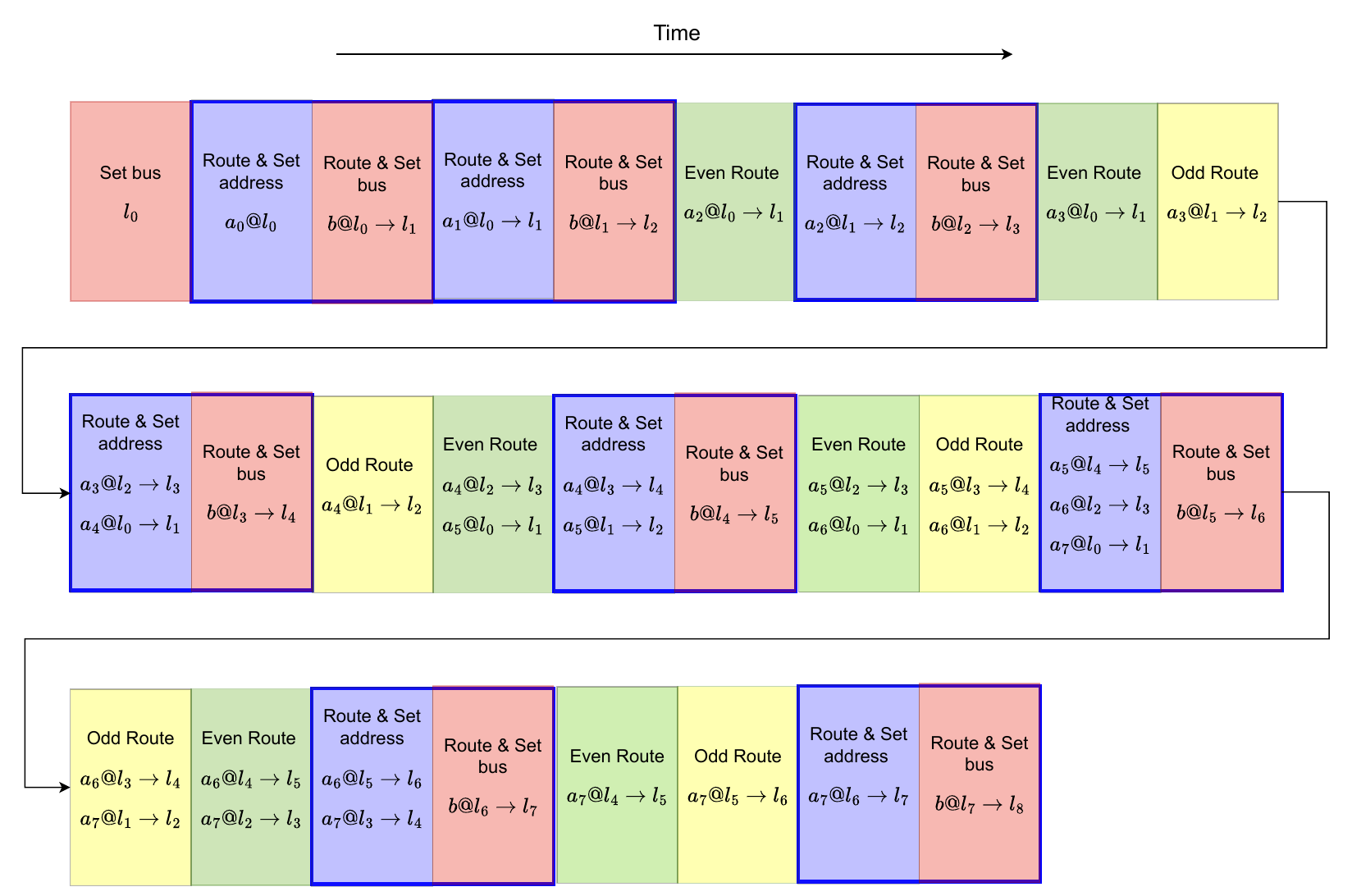}
                \label{fig:query_steps_N_256}
                \caption{Query sequence of steps for $N=8$ (Top) and $N=256$ (Bottom).}
            \end{subfigure}
            \caption{(a) Query circuit for the BB-Hetero implementation. The circuit is color coded as Red: Set bus qubits, Blue: Set address qubits, Green: Route through even routers (Red colored router nodes) and Yellow: Route through odd routers (Blue colored router nodes). (b) (Top) Sequence of steps of query circuit in (a). Colors refer to the same steps as in (a). (b) (Bottom) Sequence of steps of querying QRAM with $N=256$. }
            \label{fig:query_circuits}
            \end{figure*}

            Fig. \ref{fig:query_circuits}(a) shows the query circuit for the BB-inspired heterogeneous architecture with $N=8$. The figure is color coded to represent the different steps involved in performing the query. To simplify understanding the circuit, \ref{fig:query_circuits}(b)(Top) shows the sequence of steps of the same circuit. As evident from the figure, each router is made of three qubits (qutrits): data, address, and bus. As the address bits are routed (green and yellow steps) and set (blue steps) in the corresponding routers, the bus qubit is also routed through the tree (red steps). As a result, rather than routing the bus qubit through the tree at the end of the address setting step (as is done in \cite{hann2021resilience}), it is routed along with the address bits to prevent the $O(n)$ bus routing overhead.  
        
            Fig. \ref{fig:query_circuits}(b) (Bottom) shows the sequence of steps for a larger memory with $N=256$. In this sequence, we can see the parallelism based on the base circuit architecture from \cite{hann2021resilience} in routing address bits through mutually exclusive sets of qubits. For example, we see in Fig. \ref{fig:query_circuits}(a) that the address bit $a_3$ is still being routed and set while $a_4$ has been routed into the tree at level $l_0$ since both operations use a mutually exclusive set of routers for performing the routes and hence, commute. This parallelism ensures that the depth of the circuit from any level with code distance $d$ is linear in $d$. However, the routing steps being performed in parallel implies that several routing steps require $O(n)$ error correction code cycles rather than the $O(d)$ cycles as in the architecture in Sec. \ref{sec:implementation}. We will see the effect of this on error in the next section. 
            
            \subsubsection{Error analysis}
            Analyzing the sequence of steps in Fig. \ref{fig:query_circuits} (b), and for simplicity, assuming most operations have code distance $n$ a router at level $i$ need to be coherent for a conservative estimate,
            \begin{equation}
             \begin{split}
                 T_i &= 2nc(1+4(n-i))\\
                 &=2nc(1+4(d-1)),
             \end{split}   
            \end{equation}
            where $d$ is the code distance at level $i$ and $nc$ is the amount of time needed to perform a CSWAP operation, error corrected with a distance $n$ surface code. As a note, this is a conservative estimate because we consider all the operations to be of code distance $n$, which simplifies the analysis and gives us a loose upper bound on the error scaling for the architecture. We can substitute this $T_i$ into Eqn. \ref{eqn:hetero_fidelity} to get an asymptotic upper bound for the query infidelity of
            \begin{equation}
                \begin{split}
                1-F&\leq 4\epsilon'\left(2nc\sum_{d=1}^{n+1}p^{d_e}(4d-3)\right),\\
            (n\to\infty)&\leq8\epsilon'nc\frac{\sqrt{p}(1+3\sqrt{p})}{(1-\sqrt{p})^2},
                \label{eqn:scheme1_error}
            \end{split}
            \end{equation}
            which is $O(n)$ in query infidelity. Although asymptotically the query infidelity grows linearly in $n$, this is a conservative upper bound to the error scaling. It is attributed only to the error correction code cycles required for the address loading steps and not to the number of operations themselves. Hence, practically, the error scales much more slowly. We will verify this in Sec. \ref{sec:numerical_sim}. 
        
            \subsubsection{Qubit Overhead}
            Using the same qubit overhead convention as in Sec. \ref{sec:implementation}, we can count the number of qubits required to implement the BB inspired heterogeneous implementation. Following Fig. \ref{fig:query_circuits}(a) and starting from the root node, the number of qubits vary by layers as:
            \begin{itemize}
                \item Layer $0$: Logical qubits: 3, $d=5$ 
                \item Layer $1$: Logical qubits: 6, $d=4$ 
                \item Layer $2$: Logical qubits: 12, $d=3$ 
                \item Layer $i$: Logical qubits: $3\times2^i$, $d=n-i+1$ 
            \end{itemize}
            \begin{equation}
                \therefore M_{\text{total}} = 9\sum_{i=0}^{n}2^i(n-i+1)^2
            \end{equation}
            where $M_{\text{total}}$ is the sum of physical qubits in all layers as defined above. For $n\to\infty$, $M_{\text{total}} = 108N$. Again, we improve the constant factor slightly to $82N$ in Appendix \ref{apdx:efficeint_resources}. This is a considerable decrease in the number of qubits required to implement the QRAM as compared to FT-Hetero. 

    


    \section{Numerical Simulation of Heterogeneous QRAM} \label{sec:numerical_sim}
    
    \begin{figure*}
        \includegraphics[width=\textwidth]{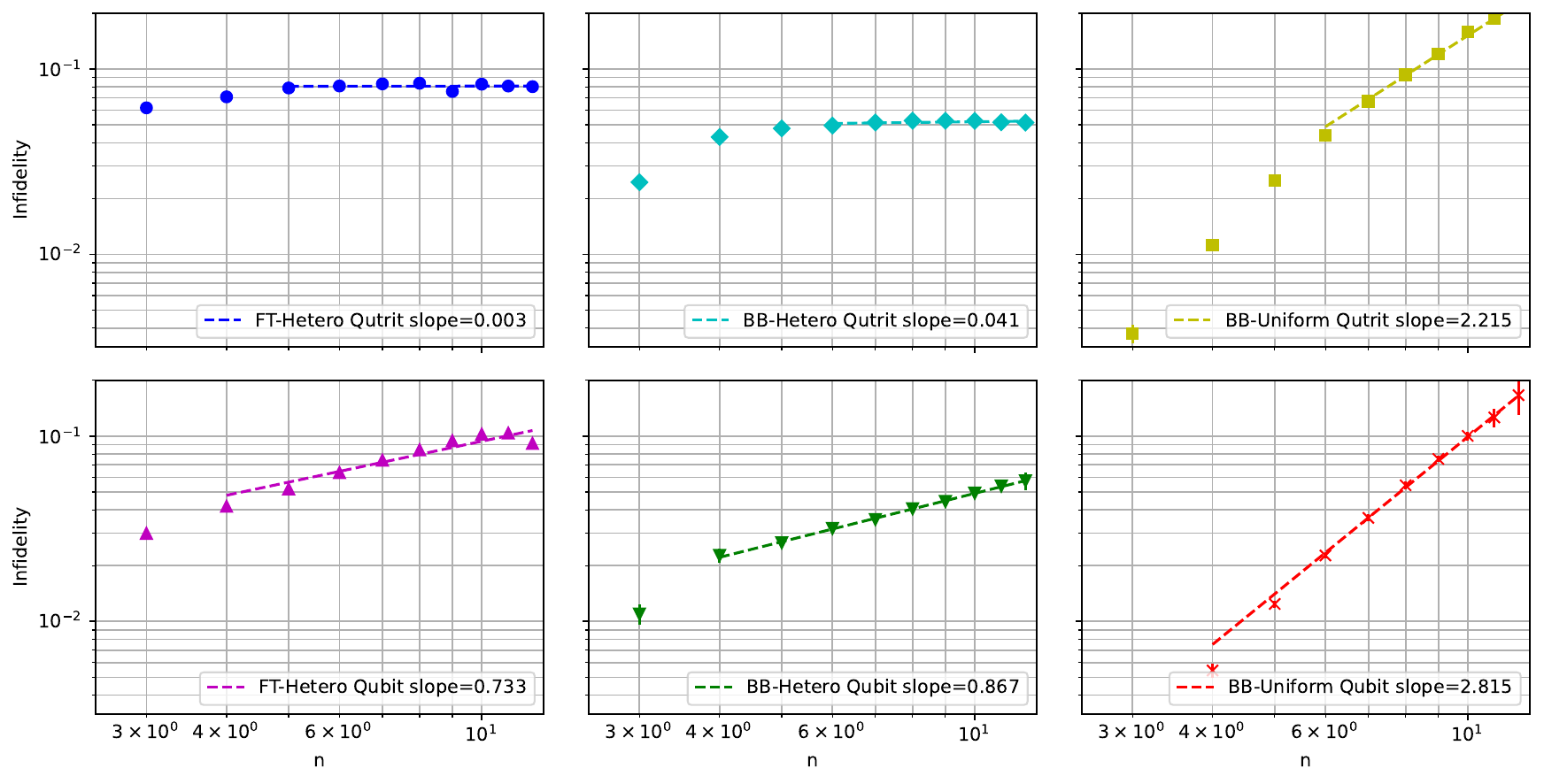}
        \caption{Numerical simulations of query infidelity as a function of QRAM circuit depth for three architectures: FT-Hetero (sec. \ref{sec:implementation}), BB-Hetero, (sec. \ref{sec:alternative_implementation}) and the Uniform BB architecture \cite{hann2021resilience}. We perform simulations for both qutrit (upper row) and qubit implementations (lower row) of the respective architectures.}
        \label{fig:aggregate_data}
    \end{figure*}
    
    We validate the Heterogeneous error scalings using numerical simulations. Since the proposed circuit QRAM architectures only use a small set of gates (SWAP and CSWAP gates), the QRAM circuits are efficiently classically simulable. This method of simulations is similar to the algorithm in \cite{hann2021resilience}. We note, however, that authors in \cite{hann2021resilience} were interested in simulating physical QRAMs. We are interested in simulating error corrected QRAMs with variable code distance surface codes. To do this, we simulate the heterogeneous architecture at the logical level, with the error rate of qubits depending on the level in which they lie. 

\subsection{Simulated Error Model}
    We use the surface code error rate described in Eqn. \ref{eqn:error_corrected} with $d = n-l+1$ where $l$ is the level of the tree in which the qubit lies. We simulate each logical qubit as a classical bit in a bit string which is operated on using bitwise operations. Quantum superpositions are modeled as a list of classical states with corresponding complex amplitudes. Coherence is assumed between all of the classical states in the list, forming a pure quantum state. We simulate open quantum system dynamics using Monte Carlo simulations where Kraus operators are sampled probabilistically from the noise model under consideration. Since we are simulating CSS codes, which project complex noise processes into $X$ and $Z$ bases using stabilizer measurements, we consider the physical error rate to be much smaller than the threshold error rate to observe exponential error suppression, and we only simulate the bit and phase flip Kraus operators. The errors are applied on all parts of the QRAM (i.e. address, bus, and other ancillary qubits) after every layer of parallel gates, which acts as gate errors on qubits being operated on and idling errors on others.

    We use the surface code with lattice surgery to simulate variable code distance patches interacting with each other \cite{fowler2018low}. The error rate assumed in the simulation is the per code cycle error rate \cite{fowler2012surface}. Hence, to simulate the error incurred due to multiple code cycles required for performing every operation, the Kraus operators are applied as many times as the largest code distance operation being performed in the parallel layer of gates. For example, if there is a root node operation being performed, all the qubits in the heterogeneous QRAM go through $n+1$ applications of the Kraus operator to simulate the $n+1$ code cycles required to error correct the root node. This simulates the effect of noise during syndrome measurements. We compare heterogeneous QRAM to a uniformly error corrected BB QRAM baseline.

\subsection{Simulation Results}
    We simulate both qubit and qutrit QRAMs. Since most quantum error correction codes are designed to work with logical qubits, we simulate qutrit quantum routers using two qubits, where the error is applied on both the qubits independently. 
    Fig. \ref{fig:aggregate_data} show the results for the FT-Hetero and BB-Hetero, and also the uniformly error corrected BB QRAM circuits. 
    
    We can see from the slopes of infidelity plots vs. QRAM depth in Fig. \ref{fig:aggregate_data} that the Heterogeneous QRAM's (both FT and BB) infidelities are upper bounded by a constant. We see that for the BB-Hetero, the $O(n)$ scaling due to the address loading steps (described in Sec. \ref{sec:alternative_implementation}) do not affect the error scaling adversely, thereby producing optimal error scaling, which is similar to the FT implementation. We can also see the $\sim O(n)$ scaling of both the qubit heterogeneous implementations. We provide the theoretical error scaling for the qubit implementations of the heterogeneous QRAMs in Appendix \ref{apdx:qubit_error_analysis}. As a baseline, we also simulate the uniformly error corrected BB QRAM and verify the $\sim O(n^2)$ and $O(n^3)$ infidelity scalings for the qutrit and qubit BB QRAMs, where we see the superior error scaling of the heterogeneous scheme against contemporary implementations. We compare the heterogeneous architectures with the uniformly error corrected architecture and other implementations in Sec. \ref{sec:comparison}.

    \section{Comparison with Prior QRAM Architectures} \label{sec:comparison}
    
    \begin{figure*}
    \centering
    \begin{subfigure}{0.5\linewidth}
        \includegraphics[width=\textwidth]{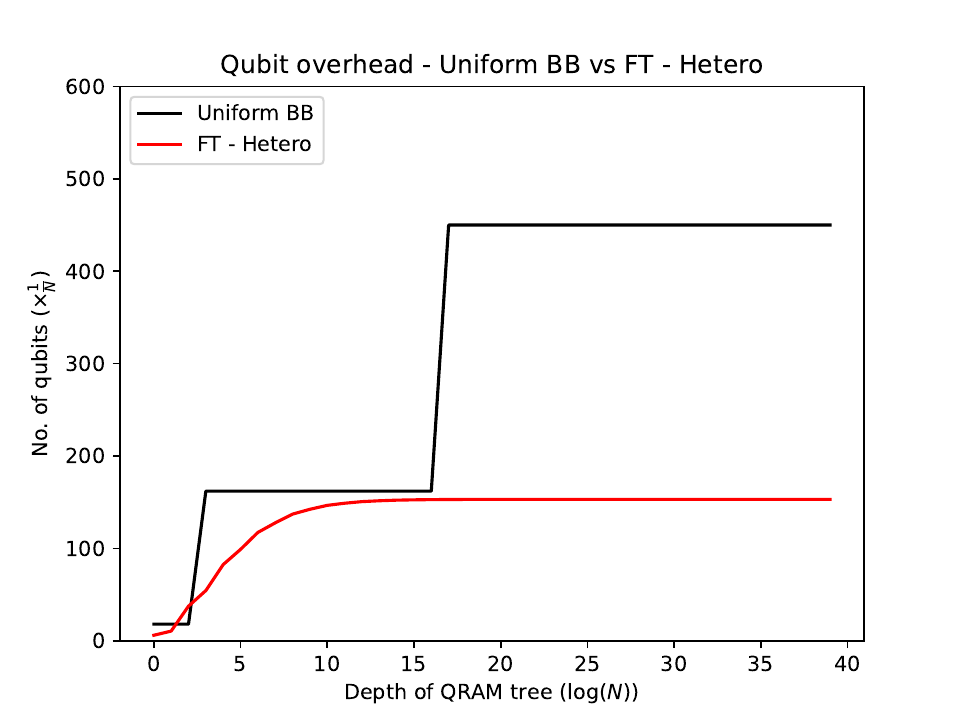}
        \caption{}
    \end{subfigure}~~
    \begin{subfigure}{0.5\linewidth}
        \includegraphics[width=\textwidth]{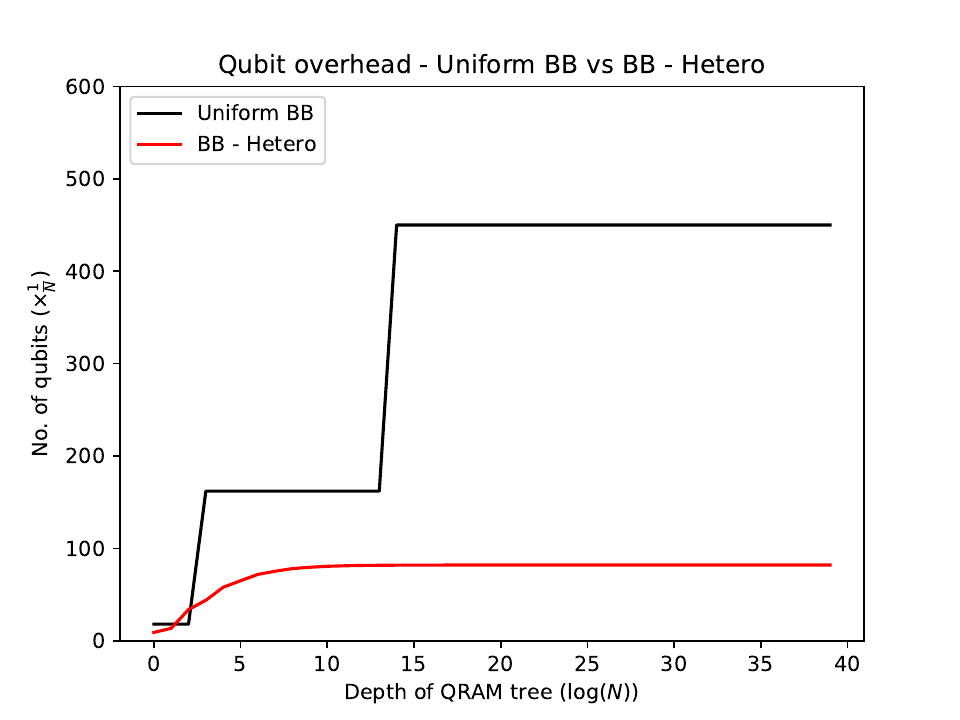}
        \caption{}
    \end{subfigure}
    \caption{Qubit overhead comparisons of uniform BB QRAM with the two heterogeneous implementations: (a) FT-Hetero and (b) BB-Hetero. The code distance for the uniform QRAM is calculated by equating its query infidelity with those of the heterogeneous implementations which is then used to calculate the number of qubits in the QRAM. The plots for the heterogeneous implementations' overheads is based on the resource scalings in Appendix \ref{apdx:efficeint_resources}.}
    \label{fig:qubit_overheads}
    \end{figure*} 

    \subsection{Resource Overhead}
    Firstly, we compare the heterogeneous QRAM against the uniformly error corrected BB QRAM in terms of the resources required to implement both QRAMs. Based on the analysis in \cite{hann2021resilience} for the BB QRAM, we can simply replace the error in Eqn. \ref{eqn:original_infidelity} with the error corresponding to the error corrected qubits with a distance $d$ surface code. Such a uniformly error corrected approach using surface codes was also proposed in \cite{xu2023systems}, although \cite{xu2023systems} suggested using rectangular patches of surface codes due to the asymmetric noise in their implementation. Hence, $\epsilon_L \rightarrow \epsilon'(p')^{d_e}$. Similarly, the number of qubits required for the $O(N)$ routers is (up to a constant factor for the number of logical qubits per router) is $M_{\text{total}} \propto 3d^2(3\times(2N-1))\approx 18d^2N$, using surface code packing efficiency $3d^2$. Using the query model in \cite{hann2021resilience} and the uniformly error corrected QRAM, the query time for the BB QRAM is:
    \begin{equation}
        T_n = 4dc(1+4n)
    \end{equation}
    which can be used to calculate the query infidelity by substituting in Eqn. \ref{eqn:original_infidelity}. Using this, we can now compare using scaling arguments the resources required to implement the heterogeneous architectures and BB QRAM. We compare the qubit overhead of the uniform BB QRAM when it produces queries with the same fidelity as the heterogeneous implementations for varying sizes of QRAMs. 
    To calculate the overheads, we find the minimum surface code distance required for the uniform BB QRAM to have at least as much fidelity as the heterogeneous numerical estimates from Sec. \ref{sec:numerical_sim}. The code distance is used to calculate the number of qubits required to implement the uniform BB QRAM and compare with the resource estimates for the heterogeneous implementations (refer to Appendix \ref{apdx:efficeint_resources}). Fig. \ref{fig:qubit_overheads} shows the calculated overheads for both the FT-Hetero and BB-Hetero implementations. Note that the qubit overhead of the uniform BB QRAM increases in steps since the surface code distance is an integral function which leads to large steps in resource overhead as the size of the QRAM increases. We can see that the heterogeneous QRAM scales considerably more efficiently as compared to the uniform BB QRAM as seen when the heterogeneous QRAM requires up to $\sim5$X fewer resources as compared to the uniform BB QRAM for $n=30$. This proves both the superior error and resource scaling of the heterogeneous scheme over the uniformly error corrected implementations.  

    \subsection{Baseline comparisons}
    There have also been other proposals for error correcting QRAM queries in literature. One of the most popular such architectures was the distillation-teleportation architecture proposed in \cite{dalzell2025distillation}. The work proposed generating physical QRAM resource states using Ballistic devices, which are then encoded onto a QEC code and distilled to generate queries of arbitrary fidelity. The main advantage of the distillation/teleportation scheme is the reduced logical resource overhead (requiring $O(\text{Poly}(n))$ logical resources as compared to $O(N)$ logical resources required for the uniformly error corrected case). However, the protocol is hamstrung by the quality of the physical QRAM device used to generate the physical QRAM resource state. The protocol requires the physical resource state to have at least $1/\text{poly}(n)$ fidelity which requires the physical devices to improve as $n$ grows. Hence, if the BB QRAM is used to implement the physical QRAM device for the distillation protocol, the physical per-component error rate must decrease asymptotically as $O(1/n^2)$ to be useful for the protocol. This limits how far the QRAM can be scaled with limited physical device coherence. Since the heterogeneous QRAM depends on variable code distance surface codes, the QRAM can be scaled to any size as long as the physical qubits are below the surface code threshold and good enough to produce the required query fidelities for much smaller QRAM sizes. 

    Another baseline comparison is against the recently introduced quantum walker (QW) \cite{de2025resource} QRAM. The QW architecture is based on the BB QRAM where the routers do not store the addresses and act as passive devices. This reduces the number of active components required to implement the QRAM from exponential active routers to just logarithmic quantum walkers, but it still requires an exponential number of passive quantum devices and two qubit operations. To analyze the QRAM, we used the simulations described in Sec. \ref{sec:numerical_sim} to perform an error analysis of the quantum walker QRAM and provide details in Appendix \ref{apdx:walker_QRAM}. 
    
    Table \ref{tab:compareQRAM} compares the two heterogeneous implementations with the uniformly error corrected BB QRAM, the distillation/teleportation protocol and the quantum walker QRAM. We can see that the heterogeneous and distillation/teleportation QRAMs demonstrate the optimal error scaling. However, the two schemes tradeoff better qubits and more qubits in their implementations. The distillation/teleportation protocol requires fewer qubits, but requires better qubits to implement the physical QRAM device as the size of the query increases as the physical component error rate must decrease as the square of the depth of the QRAM. On the contrary, the heterogeneous QRAM does not put any scaling constraints on the quality of the qubits, as long as we have enough qubits required to implement the QRAM. It is important to note that the distillation/teleportation protocol also requires an exponential amount of resources to implement the QRAM, albeit exponential physical resources, as compared to the exponential logical qubits required to implement the other QRAMs. 
    
     On the other hand, the QW QRAM requires only logarithmic resources to implement the BB architecture. The QW QRAM with the distillation/teleportation protocol would produce the most optimal resource scaling for generating fault tolerant queries. However, the QW QRAM also suffers from the quadratic error scaling of the BB QRAM since it's based on the BB architecture (see Appendix \ref{apdx:walker_QRAM}). Hence, the QW QRAM would also need to satisfy the physical device constraints levied by the distillation/teleportation protocol on the BB QRAM.
    
    Hybrid architectures made by combining multiple schemes above can be used to scale QRAMs efficiently. For example, the heterogeneous scheme can be used to supplement the distillation/teleportation scheme to scale the distillation scheme efficiently since the heterogeneous QRAM's error scaling satisfies the protocol's query fidelity requirements. 
     
    \renewcommand{\arraystretch}{1.5}
        \begin{table}
            \centering
            \begin{tabular}{|p{0.2\linewidth}|p{0.1\linewidth}|p{0.1\linewidth}|p{0.1\linewidth}|p{0.1\linewidth}|p{0.1\linewidth}|}
                \hline
                 & FT-Hetero (Our) & BB-Hetero (Our) & BB-Uniform \cite{hann2021resilience} & Distil/ teleport \cite{dalzell2025distillation} & Quantum Walker \cite{de2025resource}\\ 
                \hline\hline
                Query Time & $O(n^2)$ & $O(n)$ & $O(n)$ & $O(n)$ & $O(n)$\\
                \hline
                Num Qubits & $O(N)$ & $O(N)$ & $O(N)$ & $O(n)$ & $O(n)$\\
                \hline
                Query infidelity & $O(1)$ & $O(1)$ & $O(n^2)$ & $O(1)$ & $O(n^2)$\\
                \hline
                RPQI & $O(1)$ & $O(1)$ & $O(1)$ & $O(n^{-2})$ & $O(1)$\\
                \hline
            \end{tabular}
            \caption{Comparing QRAM architectures. RPQI represents the Required Physical Qubit Infidelity.}
            \label{tab:compareQRAM}
        \end{table}
    \section{Discussion and Conclusion}\label{sec:conclusion}

    In this work, we introduce a heterogeneous QRAM architecture using variable strength surface codes. The main motivation for the heterogeneous QRAM is the favorable error and resource scaling as compared to contemporary fault tolerant architectures. Error correcting QRAMs is an area of active research and proposals often try to tradeoff query fidelity vs. resource requirements. The heterogeneous scheme takes one step towards striking that balance. The optimal error scaling and efficient resource utilization make it ideally suited for future fault tolerant applications.

    Considerable future work will be inspired by the heterogeneous proposal. For example, the heterogeneous QRAM adopts the surface code for the reasons in Sec.~\ref{sec:surface_codes}, and notable improvements in infidelity and qubit overhead are observed. However, other QEC codes could be used, which may lead to better resource utilization. Furthermore, for our resource estimations, we consider a general packing efficiency of $3d^2$ (see Sec. \ref{sec:implementation}). Several proposals have been put forward to implement QRAMs on 2D lattices \cite{zhu2024unified, xu2023systems} which use H-Tree recursion to map the tree structure of QRAMs onto a sparsely connected grid of qubits. These proposals can be modified to accommodate the heterogeneous QRAM's varying code distance, improving qubit packing efficiency.  

    

    \begin{appendices}
        \section{Error Analysis of Qubit Router architecture}\label{apdx:qubit_error_analysis}

        \begin{figure*}
            \centering
            \includegraphics[width=\linewidth]{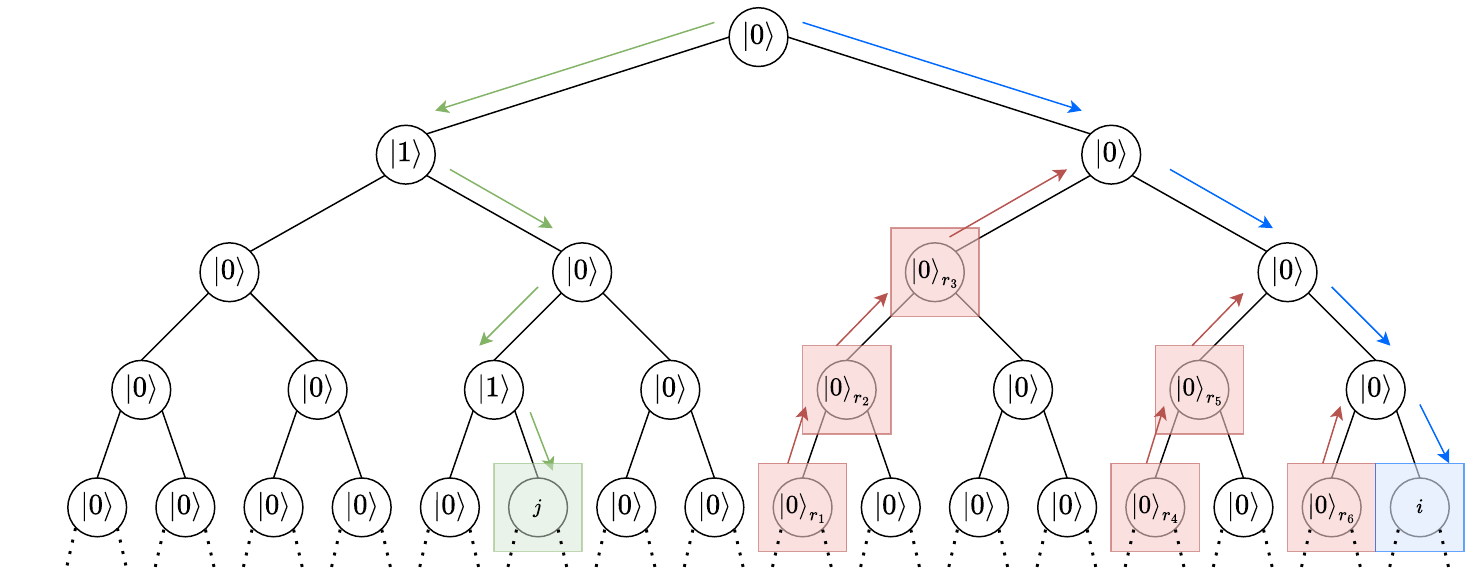}
            \caption{Error propagation from bad branches (routers marked red) may propogate into good (unqueried) branches ($i$) while another branch is being queried ($j$). Errors generated in the red blocks can propagate into the branch marked by blue arrows which hurts the query fidelity.}
            \label{fig:qubit_BB_QRAM}
        \end{figure*}
        
        In sec. \ref{sec:error_analysis}, we performed an error analysis of the proposed QRAM when the routers were considered to have three possible states - $\ket{0}, \ket{1}$ and $\ket{W}$ where the third state is the wait state. By default, the Wait state does not route the qubits to either of the children, thereby constraining the error \cite{hann2021resilience}. This, however, is not the case when the routers are qubits. In this case, the routers are always in the $\ket{0}$ or $\ket{1}$ states, due to which errors can propagate into other branches. Eventually, this leads to a query infidelity of the order of $O(\log^3(N))$ in the uncorrected case. We now analyze the query infidelity for the heterogeneous implemented with the qubit architecture.  
    
        To understand the proof of calculating query infidelity, we first show how \cite{hann2021resilience} came to the $O(\log^3(N))$ result. The proof for the heterogeneous implementation would follow similarly. The outline of the proof from \cite{hann2021resilience} is as follows: since the noise in the qubit router architectures are not constrained, the errors can propagate into good branches. More concretely, for $i,j\in g(c)$, where $g(c)$ is the set of good, i.e. error free, branches in configuration $c$, $i$ may suffer from errors propagated into it from bad branches, when the branch $j$ is queried. We note that error from bad branches cannot propagate into the queried branch but, they may propagate into good branches that are not being queried, or are being queried in superposition, reducing the query fidelity. The expected number of good branches (i.e. $g'(c)$) hence reduces by a factor: $\mathbb{E}(\Lambda') = (1-\delta)\mathbb{E}(\Lambda)$ where (to leading order) $\delta$ is defined by:
        \begin{equation}
            \delta = \epsilon\sum_{r,t}P_{r\rightarrow i} + O(\epsilon^2)
        \end{equation}
        where $r$ represents the set of routers which can propagate errors into the branch $i$. There are $O(n^2)$ such nodes which can propagate errors into the branch $i$ (errors from left edges can propagate into routers initialized as $\ket{0}$ state). Fig. \ref{fig:qubit_BB_QRAM} represents a specific example of queried branch $j$ and set of routers \{r\} which can propagate errors into the good branch $i$. We can see from this example that the number of routers in this set is of the order $O(n^2)$. For making a conservative scaling argument, they assume $P_{r\rightarrow i} = 1$, giving us the result: $\delta = \epsilon Tn^2$.
    
        Our QRAM differs from the BB QRAM in that the different levels of the QRAM have different errors. As a result,
        \begin{equation}
            \delta = \sum_{r,t}\epsilon_rP_{r\rightarrow i} + O(\epsilon^2).
        \end{equation}
        Following the example in Fig. \ref{fig:qubit_BB_QRAM}, we take $P_{r\rightarrow i} = 1$ as a conservative estimate. Now, for BB-Hetero, substituting $\epsilon_r$ as in Eqn. \ref{eqn:error_corrected}, $\delta$ can be written (to the leading order) as:
        \begin{equation}
        \begin{split}
            \delta &=\sum_{d=1}^{n+1}(n-d)(\epsilon_d)T_d\\
            &=n\sum_{d=1}^{n+1}(\epsilon'p'^{d_e})O(d^2) - \sum_{d=1}^{n+1}(\epsilon'p'^{d_e})O(d^3)\\
            &= O(n)
        \end{split}
        \end{equation}
        where we got the initial expression for $\delta$ since a level with code distance $d$ has $n-d$ routers which can propagate errors into the branch $i$, each of which has an error rate $\epsilon_d$ and maintains coherence for time $T_d$. Note that although we derived this for the specific example in Fig. \ref{fig:qubit_BB_QRAM} this scaling applies more generally in the worst case complexity. 
    
        For FT-Hetero, an error in any one of the address bits would propagate errors into the branch $j$. So, the probability of any address qubits facing an error is (to leading order):
        \begin{equation}
            \begin{split}
            \epsilon''_d &= 1-(1-\epsilon_d)^{d-1}\\
            &=(d-1)\epsilon_d
        \end{split}
        \end{equation}
        where $\epsilon_d$ is defined as Eqn. \ref{eqn:error_corrected}. Making the substitutions:
        \begin{equation}
            \begin{split}
                \delta &=\sum_{d}(n-d)(\epsilon_d)T_d\\
                &= n\sum_{d}T_d(d-1)\epsilon_d - \sum_{d}dT_d(d-1)\epsilon_d \\
                &= n\epsilon'\sum_{d}O(d^4)p'^{d_e} - \epsilon'\sum_{d}O(d^5)p'^{d_e} 
            \end{split}
        \end{equation}
        where $T_d=O(d^3)$ for the FT-Hetero. Hence, $\delta = O(n)$. 

    \begin{figure}
        \centering
        \includegraphics[width=0.75\linewidth]{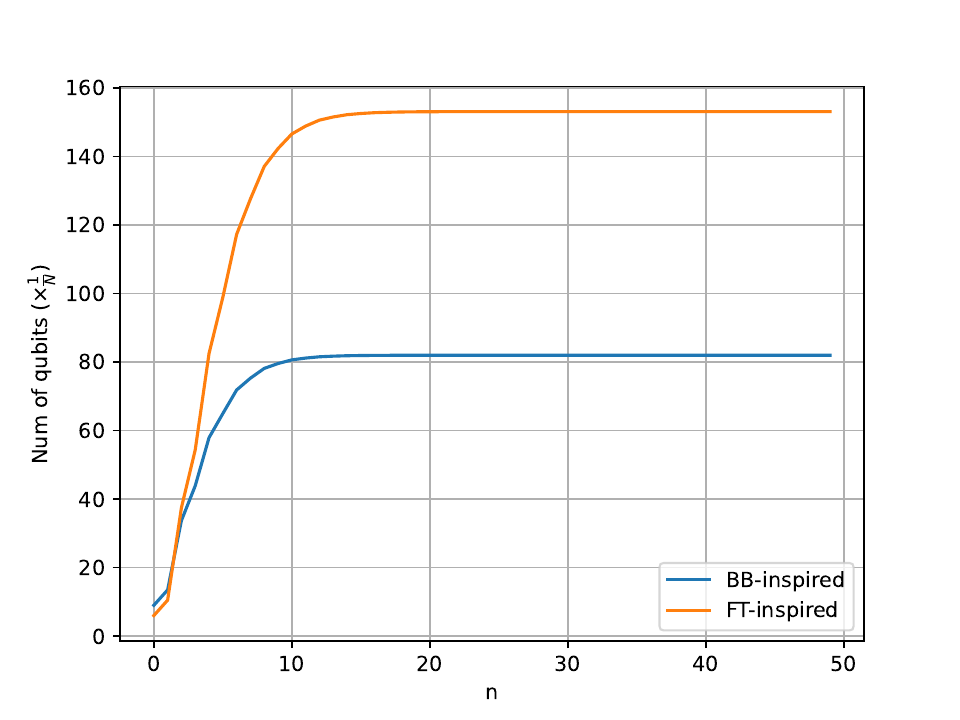}
        \caption{Resource overhead scaling using the modified qubit distribution.}
        \label{fig:efficient_VarSEQ}
    \end{figure}

    \section{Increasing Qubit Efficiency of the Heterogeneous QRAM} \label{apdx:efficeint_resources}
    We can improve the constant factor in the number of qubits required to implement both the heterogeneous implementations in Sec. \ref{sec:implementation} and \ref{sec:alternative_implementation}. This is based on the observation that the surface code error rate is given by Eqn. \ref{eqn:error_corrected}. From Eqn. \ref{eqn:error_corrected}, we can observe that for every odd code distance, $2d-1$, the next code distance $2d$ has the same error rate. Hence, instead of the code distance being distributed as 1,2,3,4,5,6..., if its distributed as 1,1,3,3,5,5,..., the fidelity of the QRAM query would not change due to the similar surface code error rates (all the error rates from sec. \ref{sec:architecture} would still apply), but the QRAM would require fewer resources. Using this new distribution, we estimate the resource requirement for the two heterogeneous implementations numerically. To do this, we calculate the number of qubits in increasingly larger sized QRAMs for both heterogeneous implementations and plot them in fig. \ref{fig:efficient_VarSEQ}. 
    
    From the plots, we can see that the qubit counts for the two implementations asymptote at 82 for BB-Hetero and $\sim$153 for FT-Hetero. These are $\sim$20\% decreases in the resources required from the default heterogeneous implementations. 
    
    \section{Quantum walker QRAM analysis} 
    \label{apdx:walker_QRAM}

        \begin{figure}
            \includegraphics[width=\linewidth]{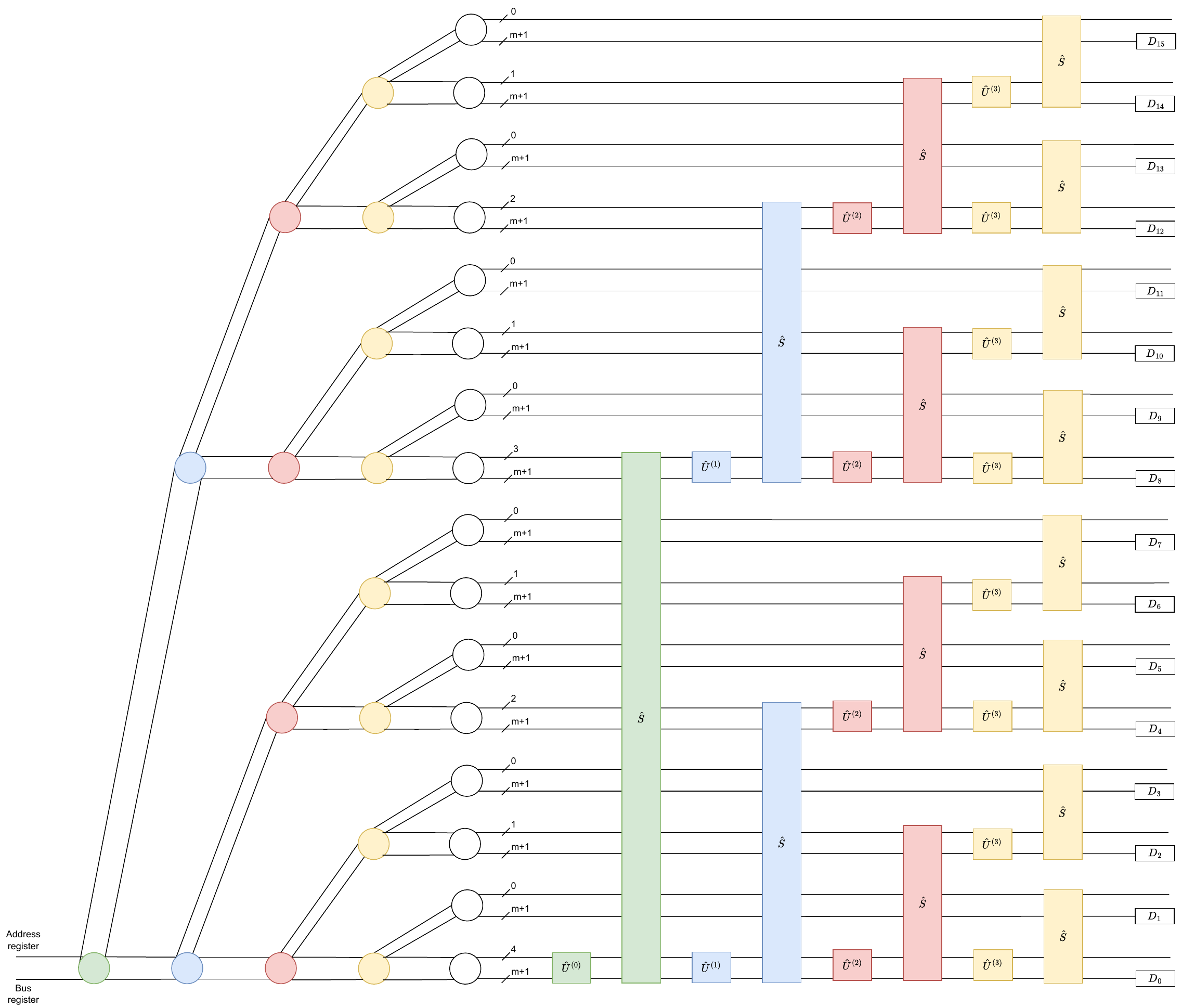}
            \caption{Quantum walker QRAM circuit. The nodes in the tree and gates in the circuit are color coded to represent different levels of the QRAM tree.}
            \label{fig:walker_circuit}
        \end{figure}

        \begin{figure}
            \includegraphics[width=\linewidth]{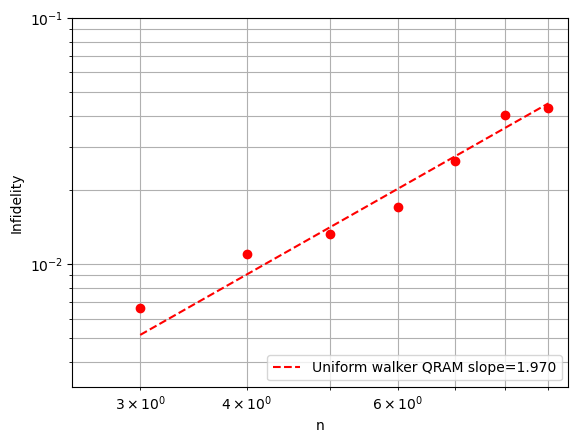}
            \caption{Numerical simulation results for the QW QRAM. }
            \label{fig:walker_data}
        \end{figure}
    
    We used the simulations described in sec. \ref{sec:numerical_sim} to analyze the error scaling of the quantum walker QRAM described in \cite{de2025resource}. Although the quantum walker QRAM was meant for bosonic systems, the authors also provided a generalization to fermoionic systems using qudits. We use two qubits to represent a qutrit system with the states $\ket{B}, \ket{R}$ and $\ket{\phi}$ with the same operations $\hat{U}^{(i)}$ and $\hat{S}$. The operator $\hat{U}^{(i)}$ is modeled as in appendix B1 in \cite{de2025resource}. We implement the $\ket{S}$ operator as: 
    \begin{equation}
        \begin{split}
             \hat{S} =~&\mathbb{I}-\ket{\phi, B^{d,l}}\bra{\phi, B^{d,l}}-\ket{R^{d+1, 2l},\phi}\bra{R^{d+1, 2l},\phi}  
             \\ + & \ket{R^{d+1, 2l}, \phi}\bra{\phi, B^{d,l}} + \ket{\phi, B^{d,l}}\bra{R^{d+1, 2l}, \phi}
        \end{split}
    \end{equation}  
    We model temporal modes in a spatial mode of the QRAM tree using individual individual qutrits (represented by two qubits). The corresponding quantum circuit for a depth 3 QRAM is shown in fig. \ref{fig:walker_circuit} where the nodes and gates are color coded according to the tree level. Errors in the qubits are modeled using circuit level noise models (eg. depolarizing noise). Obviously, performing the simulation as such does not demonstrate the logarithmic reaource overhead of the QRAM since in this implementation, each spatio-temporal mode of the QRAM is represented by an individual qubit (which grows exponentially), but the noise model is still valid since the qubits would still suffer from the same set of errors. 
    
    Performing the QRAM simulations for multiple sizes, we see from Fig. \ref{fig:walker_data} that the query infidelity grows quadratically with the depth of the QRAM. This is the same error scaling as the BB protocols since the quantum walker QRAM is also based on the BB QRAM architecture but the routers do not store the addresses and they propagate through the QRAM. In terms of the simulations, an address propagating through the same spatio-temporal mode is analogous to storage of the address qubit in time. Hence, we get the same error scaling as the BB architecture.
    
    \end{appendices}
    
    \bibliographystyle{IEEEtran}
    \bibliography{bib.bib}
\end{document}